\documentclass{article}
\usepackage{spconf,amsmath,graphicx}

\usepackage{multirow}
\usepackage{array}
\usepackage{hhline}
\usepackage{ctable}
\usepackage{svg}
\usepackage{arydshln}
\usepackage{booktabs}
\usepackage{hyperref}

\newcommand\fname[1]{``{\small \texttt{#1}}"}

\title{Multilingual Audio Captioning using machine translated data}
\name{Mat\'eo Cousin,
Étienne Labbé,
Thomas Pellegrini}
\address{
    IRIT, Université Paul Sabatier, CNRS, Toulouse, France \\
    \{etienne.labbe,thomas.pellegrini\}@irit.fr \\
}

\begin{document}

\renewcommand{\arraystretch}{1.3}
%
\maketitle
%
\begin{abstract}
Automated Audio Captioning (AAC) systems attempt to generate a natural language sentence, a caption, that describes the content of an audio recording, in terms of sound events. Existing datasets provide audio-caption pairs, with captions written in English only. In this work, we explore multilingual AAC, using machine translated captions. We translated automatically two prominent AAC datasets, AudioCaps and Clotho, from English to French, German and Spanish. We trained and evaluated monolingual systems in the four languages, on AudioCaps and Clotho. In all cases, the models achieved similar performance, about 75\% CIDEr on AudioCaps and 43\% on Clotho. In French, we acquired manual captions of the AudioCaps eval subset. The French system, trained on the machine translated version of AudioCaps, achieved significantly better results on the manual eval subset, compared to the English system for which we automatically translated the outputs to French. This advocates in favor of building systems in a target language instead of simply translating to a target language the English captions from the English system. Finally, we built a multilingual model, which achieved results in each language comparable to each monolingual system, while using much less parameters than using a collection of monolingual systems.
\end{abstract}
\section{Introduction}

Automated Audio Captioning (AAC) seeks to develop systems with the capability to provide textual descriptions of audio recordings. While most available datasets exclusively offer captions in English, practical applications demand the ability to generate text in multiple languages. Our study is motivated by two fundamental questions: i) Can machine-translated captions effectively be used to build AAC systems in languages other than English? ii) Is it advantageous to construct new systems as opposed to mere translation of captions generated by an English-based system?

To try to address these questions, we investigate the use of machine translation to train and evaluate systems in French, German and Spanish, using two prominent datasets in AAC: AudioCaps (AC)~\cite{kim_etal_2019_audiocaps} and Clotho (CL)~\cite{drossos_clotho_2019}. We develop monolingual models, building upon our recent ConvNeXt-Transformer architecture~\cite{labbe_iritups_2023}. Furthermore, we enhance this architecture to introduce a straightforward multilingual approach, capable of generating captions in all four languages with accuracy akin to a collection of monolingual systems. Finally, we compare the quality of captions produced by a French system with those translated from an English model, employing a French version of the AC test set. This version includes captions crafted by a human annotator for the specific purpose of our evaluation.




\section{Monolingual systems}

\begin{figure*}[tbp]
  \begin{center}
      \includegraphics[width=0.7\linewidth]{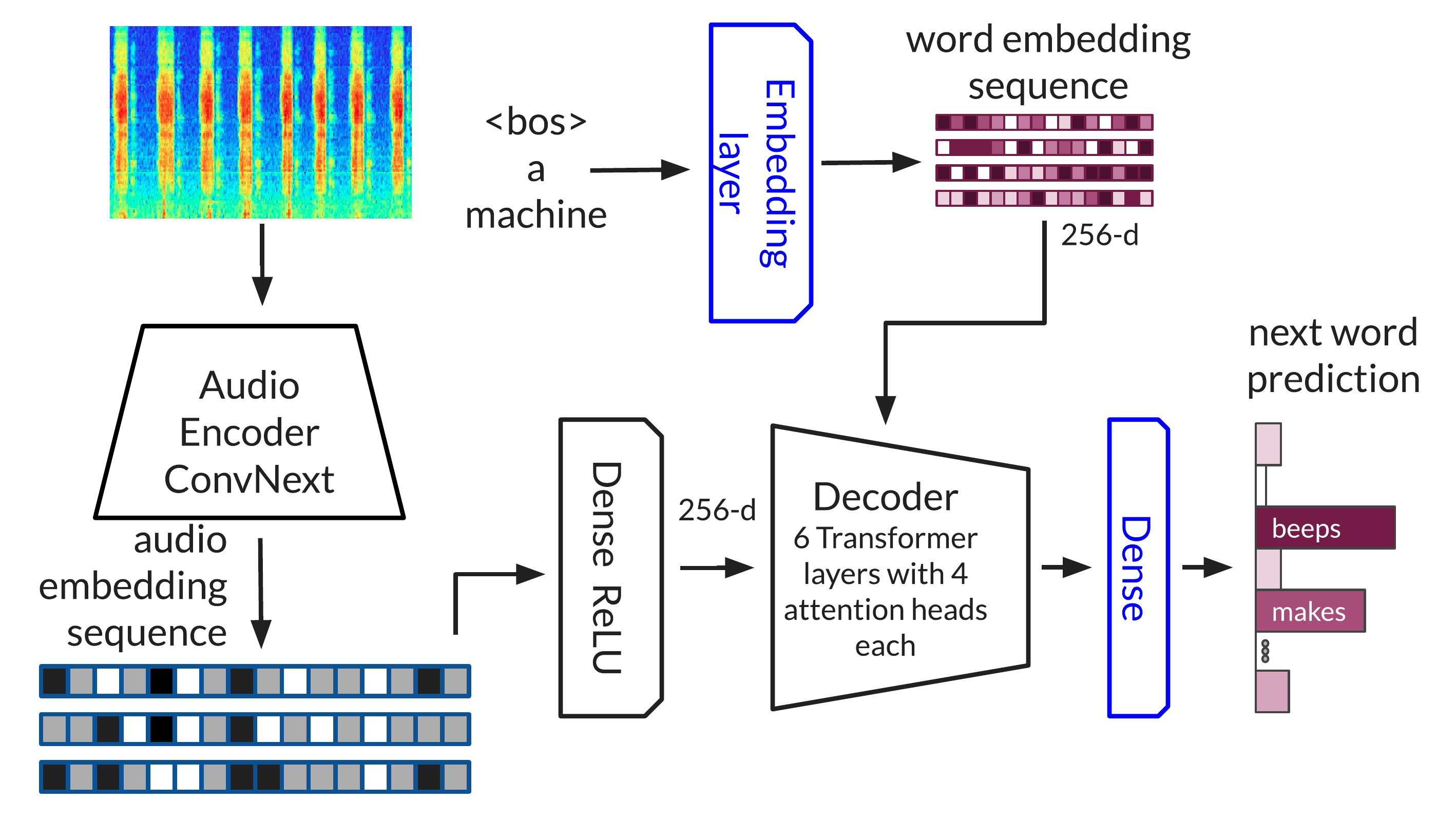}
      \caption{Overview of our AAC monolingual system: CNext-trans. Highlighted in blue are the two layers that are replicated when building a multilingual system: the embedding and classification dense layers, which are language-specific. }
      \label{fig:system}
  \end{center}
\end{figure*}

Our AAC baseline model is an encoder-decoder architecture called CNext-trans, proposed in our previous work~\cite{labbe_iritups_2023,labbe2023killing}. Figure \ref{fig:system} shows the model, comprised of an audio encoder, based on the computer vision model ConvNeXt~\cite{liu2022convnet}, which provides sequences of audio embeddings to a transformer-based language model, responsible for the word level generation of captions. 

In previous work, we adapted and pretrained the Conv\-NeXt encoder on AudioSet~\cite{audioset}, for the audio tagging task~\cite{pellegrini23_interspeech}. It achieved a 0.471 mean average precision on the AudioSet test set, better or on par with recent transformer models, such as AST~\cite{gong21b_interspeech}) and PaSST-S~\cite{koutini22_interspeech}. It gives a list of features of shape $768 \times 31$ for a 10-seconds audio clip, which are projected by a sequence of dropout set to 0.5, dense layer, a ReLU activation and another dropout set to 0.5. The decoder is a standard transformer decoder architecture~\cite{NIPS2017_3f5ee243} with six decoder layers blocks, four attention heads per block, a feedforward dimension of 2048, a GELU~\cite{gelu} activation function and a global dropout set to 0.2. Unlike many existing AAC systems, no pretrained language model was used for the decoder/word modelling part. We found that freezing the ConvNeXt encoder leads to lower variances, so we decided to precompute all its embeddings to train only the decoder part. 


For each of the four target languages, we used this CNext-trans architecture. The only architectural change needed to build a system for a new language, is adapting the size of the embedding and linear classification layers to cope with different vocabulary sizes. CNext-trans contains 28M frozen parameters (the ConvNeXt encoder) and 12M trainable parameters. This last value slightly varies according to the target language. For instance,  the classification layer is comprised of 1.2M, 1.5M and 2.3M of parameters, for English, French/Spanish, and German, respectively.




\section{Multilingual system}


We built a multilingual system based on the monolingual one, by adding language-specific embedding and classification layers in parallel, for each target language. All the remaining of the model has been kept identical. Compared to a monolingual system, with a size of about 40M parameters (12M trainable parameters, since the audio encoder is frozen), the multilingual model contains 50.8M parameters (22.8M trainable). If we were to use a collection of four monolingual systems, the total size of such systems would be 160M parameters (48M trainable parameters). Thus, the multilingual system allows to reduce the model size by 70\% relative. The main question of the present work was to compare the quality of the captions generated when using the two options.




\section{AAC Data}


AudioCaps~\cite{kim_etal_2019_audiocaps} (AC) contains 51308 audio files from AudioSet~\cite{audioset}. Since YouTube videos are sometimes removed or made unavailable, our version of the train split contains 46230 out of 49838 files, 464 out of 495 in the validation split and 912 out of 975 files in the test split.

Clotho~\cite{drossos_clotho_2019} (CL) comprises 6974 audio recordings, with duration ranging from 15 to 30 seconds. The dataset is divided into three splits used respectively for training, validation and testing, containing five captions per audio file. In our experiments, we resampled the audio files from 44.1 kHz to 32 kHz. During training, we randomly select one of five captions for each audio file.


\textbf{Machine translated versions}. AC and CL contain English-only captions. In order to translate the captions of both datasets to French, German, and Spanish, we tested two recent open-source deep learning-based tools: Opus-MT~\cite{tiedemann_opusmt_2020}, No Language Left Behind (NLLB)~\cite{nllbteam_no_2022}. A third option was considered: the paid service offered by the DeepL translation API\footnote{\url{https://www.deepl.com/translator}}. 

We compared their translations manually on a subset of sentences from AC. A number of captions in AC contain spelling mistakes, onomatopoeia and uncommon words, and we observed that both Opus-MT and NLLB failed to deliver satisfactory results in those cases. 
Given our observations and also benchmarks published in the literature~\cite{isabelle_challenge_2018,yulianto_google_2021}, we finally opted for the DeepL translation API.




\textbf{AC-fr-test-manual}. We created a manual set of captions, in French, specifically for the test subset of AC. To do so, we hired a company specialized in data annotation. A single annotator carried out the task of writing the captions. We followed the same annotation methodology as in the seminal AC study~\cite{kim_etal_2019_audiocaps}: the annotator had access to the audio track to write a caption, and in case he was not sure about the sound events he perceived, he could watch the corresponding video to eventually edit its caption. Due to budget constraints, we have limited the number of manual captions to one per sound file, instead of five in the original dataset.


In Table \ref{tab:stats}, we report the average sentence length and number of unique words in the train and test subsets in the four languages. We observe expected findings: German (de) has the smallest sentence length, whilst using much more unique words; French sentences are the longest ones; and the English vocabulary is the smallest one. 



Regarding AC-fr-test-manual, the captions are shorter than the translated ones: 11.3 words on average compared to 12.7. They use 927 unique words, which seems much smaller than the translated subset (1973 word types), but this is due to having a single caption per audio file in the manual set instead of five in the translated set.

\begin{table}[t]
\caption{Average sentence length and number of unique words in the train and test subsets of AC and CL.}
\begin{center}
\begin{tabular}{cccc}
\toprule
                            &    & Sent. length & \#Word types \\
                            &    & Train / Test     & Train / Test \\\midrule
\multirow{4}{*}{AC} & en & \hphantom{1}8.7 / 10.3                  & 4861 / 1623      \\
                            & fr & 10.4 / 12.7                     & 5797 / 1973     \\
                            & es & \hphantom{1}9.5 / 11.6          & 5889 / 1976      \\
                            & de & \hphantom{1}8.5  / 10.1         & 9391 / 2792     \\ \cdashline{2-4} 
AC-fr-test-manual           & fr & \hphantom{11}\_\hphantom{1} / 11.3                       & \hphantom{11}\_\hphantom{1} / \hphantom{1}927 \\ \midrule
\multirow{4}{*}{CL}    & en & 11.3 / 11.3                  & 4366 / 3517 \\
                       & fr & 12.9 / 12.8                  & 6384 / 3942\\
                       & es & 11.6 / 11.6                  & 6635 / 3962\\
                       & de & 10.6 / 10.6                  & 9537 / 5069\\ 
                       
\bottomrule
\end{tabular}
\label{tab:stats}
\end{center}
\end{table}



As a side-note, the total costs involved with automatically translating both AC and CL in three target languages amounted to about 300 euros, whereas the manual captions in French of the test subset of AC alone (one caption per audio) costed around 3,500 euros.

\section{Experimental setup}

To train our systems, we used the same hyperparameters as in our previous work~\cite{labbe2023killing}. The models were trained for 100 epochs on AC and 400 on CL, with the AdamW optimizer~\cite{adamw}, a $5 \cdot 10^{-4}$ learning rate, a large weight decay value of 2 and a cosine learning rate decay. We used Mixup~\cite{zhang2018mixup}  applied to the audio and word embeddings, label smoothing~\cite{szegedy2015rethinking} and SpecAugment~\cite{Park_2019}. For the multilingual systems, a training epoch consists in using each audio file four times, in order to train the models in the four languages. At inference time, we used a beam search algorithm modified to prevent the models from repeating a same word twice in a caption, except for language-specific lists of stopwords from NLTK.


To evaluate and compare the models, we report the standard metric called CIDEr-D~\cite{vedantam_cider_2015}. 
We also report semantic similarity values between the predicted captions and the reference captions in each language, by using the multilingual model {paraphrase-multilingual-mpnet-base-v2}. We will refer to this metric as SBERT-sim. 
For both metrics, the higher, the better.

\begin{table*}
\caption{Results in CIDEr-D (\%) and SBERT-sim (\%) on the test subsets of AC and CL in English (en) and French (fr), Spanish (sp), German (de), using the machine translated captions for training and testing in the case of fr, sp, de.} 
\begin{center}
\begin{tabular}{llcccccc}
\toprule
Test & \multirow{2}{*}{System} & \multirow{2}{*}{Language} & \#params ($\downarrow$) & \multicolumn{2}{c}{CIDEr-D ($\uparrow$)} & \multicolumn{2}{c}{SBERT-sim ($\uparrow$)} \\ 
data & & & mono & mono & multi & mono & multi \\

\midrule
\multirow{8}{*}{AC}
& HTSAT-BART~\cite{mei2023WavCaps} (uses external training data) & en & 171M & {78.7} & N/A & N/A & N/A \\
& Multi-TTA~\cite{kim2023exploring} & en & 108M & 76.9 & N/A & N/A & N/A \\
\cdashline{2-8}

& & en & \hphantom{1}40M & 76.1  & 77.4  & 72.2  & 72.5 \\
& & fr & \hphantom{1}41M & 79.9  & 82.0  & 75.0  & 75.1 \\
& CNext-trans (ours) & es & \hphantom{1}41M & 75.0  & 74.0  & 73.1 & 73.1\\
& & de & \hphantom{1}42M & 68.4 & 69.9 & 73.5 & 73.8\\ 
& & Avg. & \hphantom{1}41M & 74.8 & 75.8 & 73.4 & 73.6 \\

\midrule
\multirow{8}{*}{CL}
& BEATs+Conformer~\cite{wu2023_t6a} (uses external training data) & en & 1.5B & {50.6} & N/A & N/A & N/A \\
& CNN14-trans~\cite{won2021_t6} & en & 88M & 44.1 & N/A & N/A & N/A \\
\cdashline{2-8}

& & en & 40M & 46.6  & 45.6  & 58.8  & 58.5 \\
& & fr & 41M & 41.8 & 41.3  & 63.1 & 63.1 \\
& CNext-trans (ours) & es & 41M & 47.3  & 44.6 & 61.0 & 60.6 \\
& & de & 43M & 39.3  & 38.0& 64.0 & 64.1 \\ 
& & Avg. & 41M & 43.8  & 42.4  & 61.7 & 61.6\\
\bottomrule
\end{tabular}
\label{tab:results1}
\end{center}
\end{table*}

\section{Results using translated captions}

Table~\ref{tab:results1} reports the CIDEr-D (\%) and SBERT-sim (\%) values on AC and CL, achieved by monolingual and multilingual systems. In each configuration, five runs were carried out, with five different random seeds. The reported values were averaged over the five runs. Standard deviation values are not reported but they are below 1.5\% in CIDEr and 0.2\% in SBERT-sim. 

On AC, our English monolingual system achieved 76.1\% CIDEr-D, slightly worse than the state-of-the-art Multi-TTA system~\cite{kim2023exploring}, which does not use external training data. Our multilingual system slightly outperformed Multi-TTA with 77.4\% CIDEr-D. On CL, both our mono- and multilingual systems outperformed CNN14-trans~\cite{won2021_t6}.

Regarding our results in the four languages, generally speaking, similar CIDEr-D and SBERT-sim values were obtained from one language to another, except in German (de), for which the  CIDEr-D values are significantly lower, for both datasets. The multilingual model was better than the monolingual one in AC, with 75.8\% CIDEr-D and 73.6\% SBERT-sim values averaged over the four languages. In CL, it is the contrary, the monolingual system is better, by 1.4\% absolute in CIDEr-D. Table~\ref{tab_ex_1} illustrates outputs by our multilingual system for two AC audio files. We chose them to show an example (top four captions in the table) in which the captions in the four languages are identical, and another one (bottom four captions), in which they differ. 

To investigate how similar are the output captions in the different languages, we computed SBERT-sim values between them. The paraphrase-multilingual-mpnet-base-v2 model allows to directly compare captions in two different languages. With the monolingual system, the average similarity values between the English system outputs and the French, Spanish and German ones, were 77.7\% and 69.4\% in AC and CL, respectively. With the multilingual system, we compare its English outputs to its French, Spanish and German ones. We obtained larger values of 78.6 \% and 71.3\% in AC and CL, which seems to show that the multilingual system tends to produce captions that are more similar from a language to another than in the case of the monolingual models. It may be explained by the weight sharing intrinsic to the multilingual system, but further study is required to investigate this point.

\begin{table}[htb]
    \small
    \centering
    \caption{Captions obtained by our multilingual model, for two AC files, \fname{Pb6MqpdX5Jw}, and \fname{JTHMXLC9YRs}.}
    \label{tab_ex_1}

    \begin{center}
    \begin{tabular}{cl}
        \hline
        {Lang.} & {Predicted captions} \\
        \hline
        en & rain falls and thunder roars in the distance \\
        fr & la pluie tombe et le tonnerre gronde au loin \\
        es & la lluvia cae con fuerza y los truenos retumban a lo lejos \\
        de & regen fällt und der donner grollt \\
        \hline
        en & a duck quacking \\
        fr & un chien halète et respire \\
        es & los patos graznan y el viento sopla \\
        de & ein hund hechelt und atmet schwer \\
        \hline
    \end{tabular}
    \end{center}
\end{table}




\section{Results on the AC manual test set}

Table~\ref{tab:results2} reports CIDEr-D values on AC-fr-test-manual, using the monolingual systems trained in French and in English. We translated the English system outputs to French, using DeepL. We also report the score of the multilingual system. The monolingual and even more the multilingual French systems significantly outperformed the English one. 

\begin{table}
\caption{Results in CIDEr-D (\%) on AC-fr-test-manual}
\begin{center}
\begin{tabular}{lccc}
\toprule
Monolingual/     & Training & Candidates        & CIDEr-D \\
multilingual                & language(s) &                 & ($\uparrow$, \%) \\
\midrule
mono & fr & fr & 93.2 \\
multi & en, fr, sp, en & fr & 95.8 \\ 
mono & en & translated to fr & 81.1 \\ 
\bottomrule
\end{tabular}
\end{center}
\label{tab:results2}
\end{table}

\section{Conclusions}

To the best of our knowledge, this work is the first one to explore multilingual audio captioning. We demonstrate the viability of automatically translating English captions into French, German, or Spanish, and using them effectively to build proficient systems in these languages. We introduced a multilingual architecture that generates captions in all four languages. A final experiment, conducted on a newly developed iteration of the AudioCaps test set containing human-authored French captions, supports the adoption of a system trained in the target language as opposed to merely translating outputs from an English-based system.

\bibliographystyle{IEEEbib}
\bibliography{strings,refs}

\end{document}